# Growth and properties of heavy fermion $CeCu_2Ge_2$ and $CeFe_2Ge_2$ thin films

Yize Stephanie Li, Mao Zheng, Brian Mulcahy, Laura H. Greene, and
James N. Eckstein

Department of Physics and Fredrick Seitz Materials Research Laboratory
University of Illinois at Urbana-Champaign
Urbana, Illinois 61801, U.S.A.

ABSTRACT

Epitaxial films of heavy fermion $CeCu_2Ge_2$ and $CeFe_2Ge_2$ are grown on $DyScO_3$ and MgO substrates using molecular beam epitaxy. The growth begins via island nucleation leading to a granular morphology. The grains grow flat with c-axis orientation after nucleating, as indicated by *in-situ* reflection high energy electron diffraction (RHEED) and *ex-situ* analysis including atomic force microscopy (AFM) and x-ray diffraction (XRD). These single phase films show similar temperature dependent transport to single crystals of the materials indicating that similar collective order occurs in the films as in single crystals.

Heavy fermion materials are strongly correlated electron systems that exhibit very large effective masses. Within the heavy fermion family, Ce-based intermetallic compounds have attracted special attention because of the rich physics resulting from the competition between Kondo effect and Rudermann-Kittel-Kasuya-Yoshida (RKKY) interaction.[1] The recent discovery of high temperature superconductivity in electron or hole doped $BaFe_2As_2$ and $SrFe_2As_2$ has stimulated new interest to revisit the Ce-based heavy fermion 122 phases which have the same crystal structure as those FeAs 122 phases.[2] In particular, epitaxial heterostructures combining superconductivity and heavy fermion physics provide an interesting model system for understanding competing ordered phases and proximity effects.[3,4] To carry out such work, a robust process for forming films of these phases must be developed.

Two well-known examples of heavy fermion Ce-based 122 phases are CeCu$_2$Ge$_2$ (CCG) and CeFe$_2$Ge$_2$ (CFG). CCG is a Kondo-lattice system which orders antiferromagnetically below 4.15 K and exhibits an unusual temperature dependence of the resistivity caused by crystal field splitting and magnetic interactions at low temperatures.[5-7] CFG has a non-magnetic low temperature ground state which exhibits metamagnetic behavior below 12 K.[8,9] The resistivity also exhibits a distinct change to a $T^{1.5}$ temperature dependence in a temperature range from around 15 K down to 2 K, and it has been suggested this is due to the effect of enhanced spin fluctuations in the metamagnetic state caused by proximity to a nearby antiferromagnetic quantum critical point.[10] Thus, the transitions observed in the resistivity curves of single crystal samples of both materials exhibit transitions caused by collective ordering and these provide a sensitive measure of film quality. The electronic specific heat coefficient in CCG and CFG are reported to be $\gamma = 220$ and 210 mJ/K$^2$ mol respectively, which indicates effective mass about 200 times the free electron value for each.[6,8] Single crystals of CCG [5-7] and CFG [8,9] compounds have been grown and studied in the past. However, the epitaxial growth of heavy fermion thin films has been an experimental challenge. There have been some reported heavy fermion epitaxial films, including CeCu$_6$,[3,4] CeCu$_2$Si$_2$,[11] UBe$_{13}$,[12] UPd$_2$Al$_3$,[13] CeCoIn$_5$,[14] and CeIn$_3$.[15]

Here we report the epitaxial growth of CCG and CFG thin films. The films are grown by molecular beam epitaxy (MBE) using flux matched codeposition of the constituent atoms.[16] *In-situ* RHEED and *ex-situ* AFM and XRD are employed to characterize the films. We also measure the transport properties of the epitaxial films and find quantitative agreement with bulk single crystals.

CCG (CFG) has a tetragonal structure with an in-plane lattice constant of $a = 4.17$ Å ($a = 4.07$ Å) and an out-of-plane lattice constant of $c = 10.21$ Å ($c = 10.48$ Å). Each unit cell of CCG and CFG consists of two molecular layers (ML) with a relative shift of one half lattice constant ($0.5a$) in both in-plane directions. The substrate used for growth is either MgO (100) or DyScO$_3$ (110). The lattice constant of cubic structured MgO is 4.21 Å, so the lattice mismatch between MgO and CCG (CFG) is about 1% (3%). DyScO$_3$ (DSO) has a perovskite structure and the lattice constant of the pseudo-cubic unit in this material is ~3.94 Å, which has a mismatch with the CCG (CFG) lattice of 6% (3%).

Growth is done in an ultra high vacuum MBE chamber with base pressure in the low $10^{-9}$ torr range. A quartz crystal monitor (QCM), located on the rotating substrate positioner at the same horizontal plane as the substrate heater, is used to measure the rate of deposition of each elemental source. The fluxes of all the elements in CCG and CFG are stable enough over a period of several hours that the pre-growth QCM measurements provide an accurate measure of the elemental deposition rate for the actual growth, typically within $\pm$ 2%. The source temperature for each element is carefully adjusted to reach the flux matching condition. During growth, each molecular layer is codeposited with rate of about 0.1 Å/s, and RHEED is used to monitor the entire growth, allowing us to follow the growth in real time. In the optimum process we found, the substrate temperature is set to 460 ~ 510 °C prior to the growth, and is increased successively to 600 ~ 685 °C during the first 30 MLs of growth. It was kept at this temperature until the end of the growth. Typically, 100 MLs (i.e. 50 unit cells) of CCG or CFG are deposited for each film, so the nominal film thickness is 510 Å for CCG and 524 Å for CFG.

Starting the growth at low substrate temperature and increasing the temperature during the growth promotes the film quality for two reasons: (1) the first few MLs of the film wet the substrate better at low temperature, and (2) the degree of crystalline perfection improves as temperature increases. Investigating other conditions for the substrate temperature during the growth, we find that the surface quality of the epitaxial films is lower when a constant substrate temperature of 600 ~ 650 ˚C is maintained during growth. Growing the films at lower temperature resulted in reduced crystallinity and degraded transport, while growth at higher temperature caused a rougher surface to emerge.

The evolution of RHEED patterns for CCG grown on a DSO substrate (Fig. 1(a)-(c)) indicates that the growth begins via island nucleation leading to a granular morphology, and subsequently the grains grow flat after nucleating. During the first several molecular layers, the diffraction pattern consists of large transmission spots caused by small epitaxial 3D grains. A systematic transition to a flat 2D surface takes place for all of these films during the first 20 ML. By the time ~ 50 ML are deposited, no transmission spots are seen and the RHEED pattern shows a specular spot shape characteristic of reflection from regions ~ 100 nm in size. CCG grown on an MgO substrate exhibits similar RHEED patterns (not shown). This observation from the RHEED pattern is in quantitative agreement with the plateau size obtained from the post growth AFM images of CCG on DSO and MgO substrates (Fig. 1). The average step size separating the plateaus is ~ 100 Å on DSO and ~ 30 Å on MgO, which is a small fraction of the overall film thickness. XRD 2θ-ω scans for CCG grown on DSO (Fig. 1(h)) and MgO (not shown) confirm that the films are almost perfectly epitaxial with c-axis orientation. The RHEED patterns, AFM images, and XRD data for CFG films are shown in Fig. 2. The CFG films have a similar growth pattern and morphology to the CCG films, and are also epitaxial.

The sharp boundaries in two crystallographic directions between flat topped grains suggest an underlying epitaxial cause, such as stacking faults occurring at the boundaries. For 122 phases there are two possible ways the film can be in registry with the substrate if the interface plane is a transition metal plane. The next layer then is Ge, with those atoms occupying sites in one half of the centered positions on top of the square transition metal lattice. Thus there are two variants possible and at junctions between them a stacking fault occurs. The same is true for growth on MgO. For both substrates, it is possible that next nearest neighbor interactions would favor one Ge lattice registration over the other, but this may be small and not effective at the growth temperature used.

A more delicate measure of the sample quality in materials exhibiting electronic conduction is the emergence of collective order at low temperatures and this can often be seen in resistance measurements. The resistivity vs. temperature, $\rho(T)$, of the films is shown in Fig. 3. For the CCG film grown on the MgO substrate (Fig. 3a), three distinct features are observed, quantitatively in agreement with what is seen in bulk single crystals[5-7]: The broad peak around 110 K is attributed to crystal field splitting of the rare earth ion, $T_{CF}$; the sharp peak around 5K is the Kondo lattice coherence temperature, $T^*$; and the sharp cusp around 4K marks the onset of antiferromagnetic order, $T_N$. The temperatures at which we measure these resistivity signatures in our thin films ($_f$) and those reported for the bulk[6] ($_b$) are: $T_{CFf}$ = 112 K and $T_{CFb}$ = 110 K; $T^*_f$ = 4.7 K and $T^*_b$ = 5.2 K; $T_{Nf}$ = 3.87 K and $T_{Nb}$ = 4.15 K. Note our films exhibit these signatures at temperatures consistently within a few percent of those reported for bulk single crystals. The near quantitative agreement of these resistive transition temperatures indicate our films are of high electronic quality, and quite comparable to single crystals.

For the CFG film grown on DSO substrate (Fig. 3b) the $\rho(T)$ behavior is also similar to what is observed in single crystals of that material.[8,9] We note that films of CFG grown on an MgO substrate (not shown) exhibit nearly identical $\rho(T)$. The resistivity decreases monotonically as the temperature is lowered, similar as what is reported for CFG single crystals. The residual resistivity ($\rho_0$) of our films is 4.5 μΩ·cm, which is close to the $\rho_0$ reported for the bulk ($\rho_0$ = 4.2

μΩ·cm in Ref. 8, and $\rho_0$ = 3.7 μΩ·cm in Ref. 9), indicating the high crystallographic quality of these films. There is a subtle but distinct transition in the $\rho(T)$ curve at ~14.9 K. This can be seen by plotting the resistivity minus the residual resistivity versus temperature as shown in the inset of Fig. 3(b). Fitting the data, we find out that resistivity has a power law dependence with $\rho(T) - \rho_0 \propto T^{1.58}$ for temperatures below 14.9 K where a cusp is seen in the curve. In Ref. 9, a $T^{1.5}$ dependence of the resistivity was observed in a temperature range from around 15 K down to 2 K, which was explained in terms of spin fluctuations near an antiferromagnetic quantum critical point in Ref. 10. The power law behavior at low temperature in our films quantitatively agrees with that reported in single crystals.

In summary, c-axis oriented epitaxial thin films of heavy fermion CCG and CFG have been grown on DSO and MgO substrates using molecular beam epitaxy. The film consists of sub-micrometer-sized flat grains epitaxially oriented with respect to the substrate. The transport properties of the films are in good agreement with those of bulk single crystals. As a demonstration of the successful growth of epitaxial Ce-based 122 phases, this work may stimulate further investigation of the high quality epitaxial growth of heavy fermion films especially in combination wth new types of superconducting films that share the same structure as Ce-based 122 phases. This will enable research on these systems to include heterostructures of different components, in particular combining heavy fermion physics with a wide range of materials systems that exhibit superconductivity.

This material is based upon work supported by the Center for Emergent Superconductivity, an Energy Frontier Research Center funded by the U.S. Department of Energy, Office of Science, Office of Basic Energy Sciences under Award Number DE-AC0298CH1088. AFM and XRD analysis were carried out in the Frederick Seitz Materials Research Laboratory Central

Facilities, University of Illinois, which are partially supported by the U.S. Department of Energy under grants DE-FG02-07ER46453 and DE-FG02-07ER46471. We thank Scott MacLaren, Mauro Sardela, and Doug Jeffers for technical support and discussion.

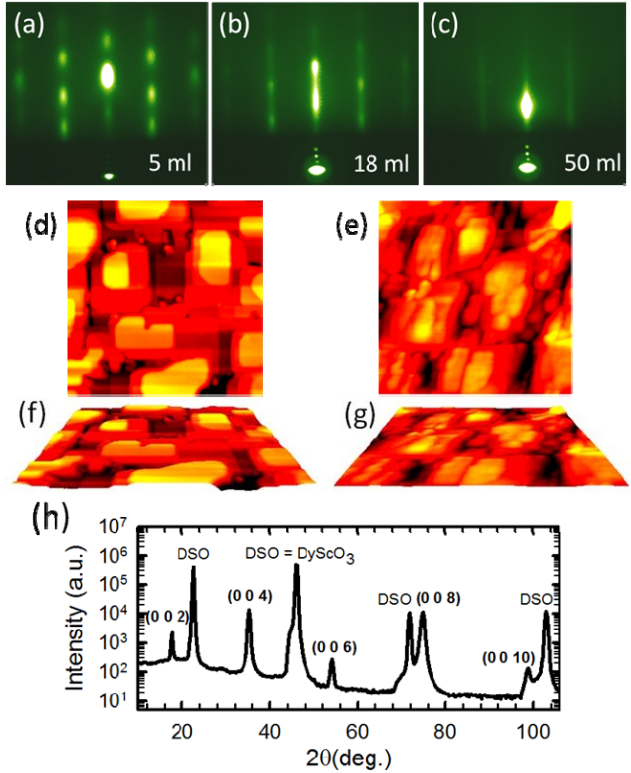

FIG. 1. (a) – (c) RHEED images of CCG film on DSO substrate at the end of 5 ml, 18 ml, and 50 ml of growth. (d) 1×1μm² AFM image of ~ 500 Å thick CCG film grown on DSO substrate, with 30 nm height scale. The typical step height size is ~ 100 Å. (e) 1×1μm² AFM image of ~ 500 Å thick CCG film grown on MgO substrate, with 10 nm height scale. The typical step height size is ~ 30 Å. (f) Perspective view of the AFM image shown in (d). (g) Perspective view of the AFM image shown in (e). (h) XRD 2θ-ω scans of CCG film on DSO substrate.

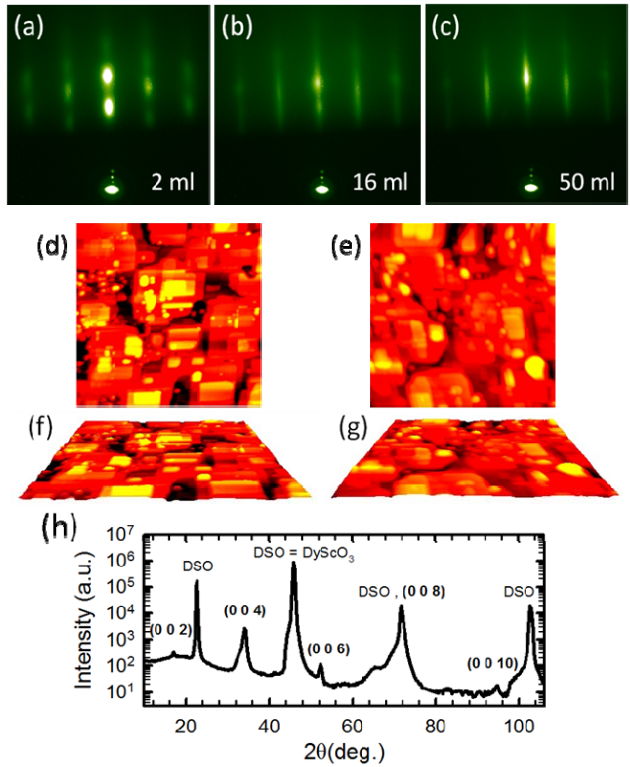

FIG. 2. (a) – (c) RHEED images of CFG film on DSO substrate at the end of 2 ml, 16 ml, and 50 ml of growth. (d) 1×1μm² AFM image of ~ 500 Å thick CFG film grown on DSO substrate, with 20 nm height scale. The typical step height size is ~ 60 Å. (e) 1×1μm² AFM image of ~ 500 Å thick CFG film grown on MgO substrate, with 20 nm height scale. The typical step height size is ~ 60 Å. (f) Perspective view of the AFM image shown in (d). (g) Perspective view of the AFM image shown in (e). (h) XRD 2θ-ω scans of CFG film on DSO substrate.

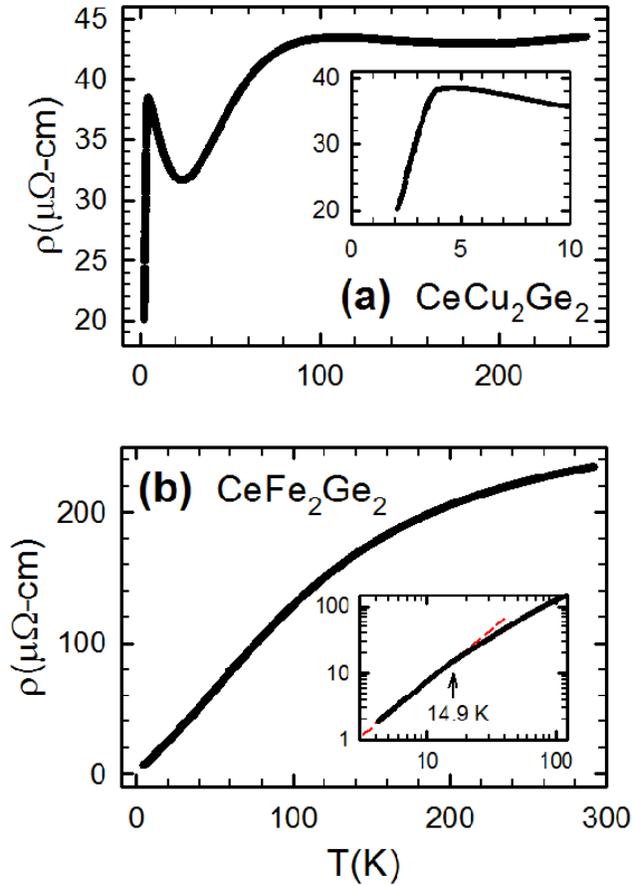

FIG. 3. Temperature dependence of resistivity for (a) CCG film grown on MgO substrate (Inset is a blow-up view of the curve below 10 K, which shown the sharp transition at ~ 4 K) and (b) CFG film grown on DSO substrate (Inset is a blow-up view of the curve below 120 K in log-log scale after subtracting the residual resistivity $\rho_0$, and the arrow indicates that the transition from $T^{1.58}$ dependence to $T^{1.20}$ dependence occurs around 14.9 K. The red dashed line is the fit to linear function for the data below 14.9 K).